\documentclass[aps,twocolumn,amsmath,amssymb,showpacs,floatfix,prb]{revtex4}
\usepackage{graphicx}

\newcommand{\etal}{{\it et al.}}
\newcommand{\beq}{\begin{equation}}
\newcommand{\eeq}{\end{equation}}
\newcommand{\bea}{\begin{eqnarray}}
\newcommand{\eea}{\end{eqnarray}}

\begin{document}
\title{Gapless superconductivity and the Fermi arc in the cuprates}
\author{A. V. Chubukov$^1$, M. R. Norman$^2$, A. J. Millis$^3$ and E. Abrahams$^4$}
\affiliation{
$^{1}$ Department of Physics, University of Wisconsin-Madison,
 Madison WI 53706-1390 \\
$^{2}$Materials Science Division, Argonne National Laboratory, Argonne,
IL 60439\\
$^3$ Department of Physics, Columbia University,  538 W. 120th St, New York, NY 10027\\ 
$^4$ Center for Materials Theory, Serin Physics Laboratory, Rutgers  
University, Piscataway, NJ 08854}

\date{\today}
\begin{abstract}
We argue that the Fermi arc observed in angle resolved photoemission 
measurements 
 in underdoped cuprates can be understood as a consequence of inelastic scattering in a d-wave
superconductor.  We analyze this phenomenon
in the context of strong coupling Eliashberg theory, deriving 
 a `single lifetime'
model for describing the temperature evolution of the spectral gap as measured by single particle
probes such as photoemission and tunneling.
\end{abstract}

\maketitle

Underdoped cuprates are characterized by a gap which
 exists in the excitation spectrum even above the superconducting transition temperature $T_c$  and
 is detectable up to a temperature $T^*$, the ``pseudogap temperature".
The origin of this pseudogap has been a source of much debate.
One of the most intriguing experimental characteristics of this pseudogap 
 regime is a
 partial destruction of the normal state Fermi surface: 
 states near the edges of the two-dimensional Brillouin zone (`antinodes')  are
  gapped below $T^*$, while states along an arc
 near the zone diagonals (`nodes')
 remain gapless.
This phenomenon 
has been extensively measured 
 in angle resolved photoemission (ARPES) experiments~\cite{arpes,norm_98,arpes_camp}.
The convention  used to interpret 
ARPES data is that a 
point  ${\bf k}_F$ on the Fermi surface is  considered to be gapped if 
 the spectral function $A({\bf k}_F, \omega) \propto |{\rm Im}\, G ({\bf k}_F, \omega)|$ 
 has a maximum at a
non-zero
 frequency  and is gapless if the 
maximum of $A({\bf k}_F, \omega)$ is located at zero frequency~\cite{norm_98}. 
 In underdoped cuprates in the temperature range between $T_c$ and $T^*$,
the gapless regions appear to form arcs rather than closed contours. 
 The most recent measurements show~\cite{arpes_camp}
 that  the length of the arc is temperature dependent: it
appears to be linear in $T$ and seems to vanish 
if the data obtained in the 
 temperature range $T_c<T<T^*$ 
are extrapolated to $T=0$.  Moreover, 
as $T$ is decreased below $T_c$, 
the arc collapses rapidly 
to a point node~\cite{latest}.  
 Understanding these Fermi arcs has become a key question in the physics of the cuprates.

Theoretical scenarios for the arc can be separated into two 
groups. One
scenario is that the arc emerges due to long range (or quasi-long range) ordering that is
not related to the pairing. The proposals include
orbital  currents~\cite{varma}, 
 SDW~\cite{joerg,tremblay}, 
 CDW~\cite{italy}, 
and dDW~\cite{morr} ordering.
 A generic consequence of these theories is the presence 
of two gaps -- a smaller superconducting gap 
in the nodal regions, and a larger gap of different origin in the antinodal regions.
Above $T_c$, the superconducting gap vanishes 
 leaving  only the other 
 order, which leads to 
Fermi surface pockets near the zone diagonals~\cite{exception}.
The ARPES intensity on one side of these pockets is expected to be
weaker than on the other side (as observed in experiments on materials with density wave ordering)
so that the pockets look like arcs in ARPES measurements.
The notion of two gaps is also implicit in scenarios relating the arc to Mott physics~\cite{gabi,phillips}.

\begin{figure}
\centerline{\includegraphics[width=3.2in]{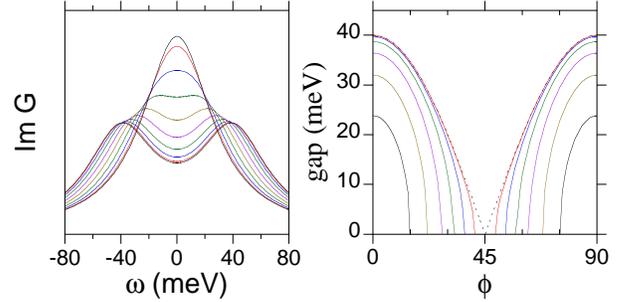}}
\caption{(a) Im G around the Fermi surface from Eq.~(\ref{1}) with 
$\Phi =\Phi_0\cos(2\phi)$,
where $\Phi_0$=40 meV and $\gamma =30 meV$.
  The top curve is at the node ($\phi=45^\circ$),
the bottom curve at the antinode ($\phi=0^\circ$), with $\phi$ increments of $5^\circ$.
(b) Spectral gap (half the peak to peak separation) versus $\phi$ for $\gamma$ ranging
from 0 (dashed curve) to 60 meV (bottom curve) with $\gamma$ increments of 10 meV.}
 \label{fig1}
\end{figure}

An alternative `one-gap' scenario interprets the pseudogap as a remnant of
the superconducting gap, with thermal fluctuations destroying long-range
order at $T \geq T_c$, but allowing gap-like features to remain in the spectra
for a range of higher temperatures \cite{Kotliar88,mohit,Emery95}.
 The effect of phase disordering on the spectral function has been
 extensively discussed in Refs.~\onlinecite{norm_98,millis_franz}. A phenomenological  
 Green's function has the form
\beq  
G ({\bf k}_F, \omega) = \frac{1}{\omega + i\gamma_2 - \frac{\Phi^2 (\phi)}{\omega + i\gamma_1}} =  \frac{\omega + i\gamma_1}{(\omega + i\gamma_1)(\omega + i\gamma_2) - \Phi^2 (\phi)}
\label{1}
\eeq
Here $\Phi (\phi)$ is the  $d-$wave anomalous self energy,
 and $\phi$ is 
 an angle
  denoting the position along the Fermi surface. 
 The scattering $\gamma_1$ comes from pairing (or vortex) fluctuations,
 and $\gamma_2$ is the fermionic scattering rate.

A simplified model with  $\gamma_1 = \gamma_2 =\gamma$  is sufficient to understand how the 
Fermi arc, as experimentally defined, emerges due to scattering. 
At small $\gamma$, ${\rm Im}\, G ({\bf k}_F, \omega)$ has sharp peaks
at $\omega = \pm\Phi (\phi)$. However, as  $\gamma$ increases, the
peaks  broaden and slowly shift to smaller absolute frequencies
(Fig.~1a).
 For $\gamma  \leq \sqrt{3} |\Phi (\phi)|$, the maxima are located at
 $\omega = \pm[2\Phi\sqrt{\gamma^2+\Phi^2}-\gamma^2-\Phi^2]^{1/2}$.
For  $\gamma > \sqrt{3} |\Phi (\phi)|$,  
a single maximum occurs at $\omega =0$.
 The boundary value $|\Phi(\phi)| = \gamma/\sqrt{3}$ 
 sets the length of the Fermi arc (Fig.~1b).
Because $\Phi (\phi)$ vanishes at the node,  the arc has a non-zero length 
for arbitrarily small $\gamma$. 

Our goal is to  provide 
 a  microscopic justification for
Eq.~(\ref{1}) (with $\gamma_1 = \gamma_2=\gamma$)
and to show that this
 is equivalent to gapless superconductivity.
 The latter 
naturally emerges at non-zero $T$ in an Eliashberg 
 theory, independent of
 the nature of the electron-boson interaction and the symmetry of the pairing gap.
Obtaining a  two-lifetime model with $\gamma_1 \neq \gamma_2$ would require additional 
physics~\cite{norm_98,millis_franz}.
We  show that 
even in the absence of impurities, $\gamma$
 is always non-zero for $T >0$.
Both $\gamma$ and the length of the arc
scale linearly with $T$ at high temperatures, when the strongest
interaction is between fermions and classical (thermal) bosons. 
 The linear $T$ behavior of $\gamma$ sets in well before the
temperature exceeds the ``effective Debye frequency'' of the bosons
for the same reason that linear $T$ resistivity
becomes apparent for $T\gtrsim \Omega_{Debye}/3$ \cite{allen}.
 The linear $T$ dependence of $\gamma$ 
 in the pseudogap phase could also arise from the same physics that gives rise
 to the marginal Fermi liquid behavior of the normal state~\cite{mfl}.

This scenario is consistent with the ARPES experiments.
  The key evidence is  the vanishing of the extrapolated arc length at $T=0$, and 
 the linear $T$ dependence of the arc length for $T > T_c$.  Consistent with this
 is the ``closing'' of the gap in the nodal region, i.e.~the reduction of the spectral peak 
position to zero
 for small $\Phi$ relative to $\gamma$, and the ``filling in'' of the gap in the 
 antinodal region, i.e.~the spectral peaks broadening without moving much for large
 $\Phi$ relative to $\gamma$.
 This behavior is evident from Eq.~(\ref{1}) (see Fig.~1a).
 For a two-gap scenario,  the arc 
length above $T_c$ should have only a weak dependence on $T$, and would extrapolate to
a  non-zero 
value in the $T$=0 limit, in contradiction to the ARPES data~\cite{exception}.

The rapid shrinking of the  arc length
 below $T_c$ 
 found in Ref.~\onlinecite{latest}
 is a more subtle issue.
Our analysis shows that a Fermi arc
with $T$-dependent 
length is a generic property
 of a $d$-wave superconductor at non-zero $T$, independent of issues associated with
 superconducting  coherence. In other words,  long-range
 superconducting order does not have to be destroyed by fluctuations for the arc to be present.  This generally implies that the arcs must survive below $T_c$.
 We find, however,
  that the length of the arc is very  small at the lowest $T$ 
(exponentially small if the spectrum of the scattering bosons is completely 
 gapped well below $T_c$), but becomes both sizable and linear in $T$ at temperatures 
 above the fluctuation-driven $T_c$.
 This would be consistent with the drop of the scattering rate 
 below $T_c$ that has
 been inferred from a variety of techniques, including ARPES and various conductivity 
 measurements~\cite{timusk}.
Another possibility, which we don't explore here, is that  the rapid variation of the arc length 
near $T_c$ is due to a strong temperature variation of the damping originating from the interaction 
between electrons and pairing and/or vortex fluctuations~\cite{norm_98,millis_franz,elihu,maki}.
 Phenomenologically this can be described as the
 vanishing of  $\gamma_1$  in the 
 two-lifetime model of Eq.~(\ref{1}).
 Once $\gamma_1 =0$ at $T_c$, $A({\bf k}_F, \omega)$ vanishes at $\omega =0$,
 and the spectral peak is located at a non-zero frequency for all ${\bf k}_F$ points except 
 the $d$-wave node~\cite{norm_98}. 
 
Gapless superconductivity 
was first discussed in the context of a  BCS $s$-wave 
superconductor with magnetic impurities~\cite{ag_gappl}.
 Although not widely appreciated, even in a clean $s$-wave system, 
the Eliashberg equations yield gapless 
superconductivity~\cite{ssw,bergmann,comb,msv}. The reason is that at non-zero $T$, the density of the 
 bosons is non-zero, and  a fermion can undergo scattering 
by on-shell thermal bosons~\cite{ssw}.  This leads to a non-zero 
damping term in the fermionic 
 self-energy $\Sigma (\omega)$, just as in a superconductor with magnetic impurities. 

To understand gapless superconductivity for
 the clean case and to discuss the Fermi arcs, 
  we consider a $d$-wave Eliashberg theory with
 an  effective (dimensionless) 
interaction $\chi ({\bf k}-{\bf k}^\prime, \Omega)$ between the electrons at the Fermi 
surface ($|{\bf k}| = |{\bf k}^\prime| = k_F$). We assume   $\chi({\bf k}-{\bf k}^\prime, \Omega)$  can be decoupled into 
 $s$-wave and $d$-wave harmonics, i.e.,
\begin{equation}
\chi ({\bf k}-{\bf k}^\prime, \Omega) = \lambda \left[\chi_d (\Omega) d_k d_{k^\prime} + \chi_s (\Omega) s_k s_{k^\prime}\right]
\label{14}
\end{equation}
where  $\lambda$ is a dimensionless coupling, $s_k = 1$, and $d_k = \frac{1}{2}(\cos k_x - \cos k_y$). We
 choose a normalization such that $\chi_s (0) =1$. 
We assume that $\chi_d$ is attractive, 
and $\chi_s$ is repulsive. 
 We also assume  for simplicity that there is no feedback effect from the pairing on $\chi 
 ({\bf k}-{\bf k}^\prime, \Omega)$ -- this last simplification may be partly justified by the fact that, 
 experimentally, feedback effects are relatively weak in  the pseudogap phase.
 
Eliashberg theory provides a mechanism for the opening of a gap in the excitation spectrum, 
without giving information on phase coherence.
 Therefore, in what follows, we shall use it to describe the pseudogap behavior for $T<T^*$.
In a superconductor, the self energy has a normal
part expressed in terms of a renormalization factor $Z_k(\omega)$ 
(which is $k$ independent if $s_k=1$),
and an anomalous (pairing) part expressed in terms
of the amplitude $\Phi_k(\omega) = d_k \Phi (\omega)$ (the equality follows from our choice of interaction, Eq.~\ref{14}). 
 The superconducting gap  is
$\Delta_k(\omega)=\Phi_k(\omega)/Z_k(\omega) = d_k \Delta (\omega)$. 
 The  quantities 
$\Delta (\omega)$ and $Z(\omega)$ 
 are given
by the solution of  two coupled integral equations~\cite{eliash}, 
of which the equation for $\Delta$ involves only $\Delta$ while the other expresses $Z$ in terms of $\Delta$.
 As long as the interaction is finite at zero frequency,
 $\Phi (\omega_m)$  tends to a non-zero value at the 
smallest $\omega_m$, 
 and for the purposes of our small $\omega$ analysis 
can be safely replaced by a frequency-independent value $\Phi$. 
 The low-frequency behavior of $Z(\omega_m)$ is more involved, and 
the conversion of $Z$ 
to real frequencies at finite $T$ is nontrivial~\cite{comb}.

Note that Eq.~(\ref{14}) is an approximation --
there is no guarantee that for a realistic  
 $\chi ({\bf k}-{\bf k}^\prime, \Omega)$  the pairing amplitude $\Phi$
will have the simple $d-$wave form $\Phi (k) = d_k \Phi$.  
 In particular, near the onset of pairing, 
the form of the gap will be such as to minimize the pairbreaking 
effects specific to systems with a non-$s$-wave gap symmetry~\cite{msv}.
 This by itself may give rise to a suppressed gap around the node, even without the 
 contribution from $Z_k (\omega)$. We, however, will not explore this possibility here 
 and simply assume $\Phi_k \propto d_k$.

It is more convenient to 
solve the gap equation directly along the real frequency axis
and to cast the Eliashberg equations in terms of  $D(\omega)=\Delta(\omega)/\omega$.  For an electron-phonon superconductor, this procedure has been exploited in
 Ref.~\onlinecite{comb} to obtain an
 equation for $\Delta (\omega)$ along the real axis.  The extension to a $d$-wave superconductor, 
 and to arbitrary  $\chi_s (\Omega)$ and $\chi_d (\Omega)$,
 is straightforward~\cite{acf} and yields 
\begin{equation}
D(\omega) B(\omega) = A(\omega) + C(\omega)
\label{3}
\end{equation} 
where
\begin{widetext}
\begin{eqnarray}
&& A(\omega) = \frac{\lambda}{2} \int_{-\infty}^\infty d \Omega 
\tanh{\frac{\Omega}{2T}} {\rm Re} \left[\frac{D(\Omega)\chi_d (\Omega-\omega)}{\sqrt{1-D^2 (\Omega)}}~\right]; ~B(\omega) = \omega + \frac{\lambda}{2} \int_{-\infty}^\infty d \Omega 
\tanh{\frac{\Omega}{2T}} {\rm Re} \left[\frac{\chi_s (\Omega-\omega)}{\sqrt{1-D^2 (\Omega)}}\right]\nonumber \\
&& C(\omega) = i \frac{\lambda}{2} \int_{-\infty}^\infty \frac{d \Omega}{\sqrt{1-D^2 (\omega + \Omega)}} \left[{\rm Im} \chi_d (\Omega) D(\omega + \Omega) -
{\rm Im} \chi_s (\Omega) D(\omega)\right]\left[\coth{\frac{\Omega}{2T}} - \tanh{\frac{\omega + \Omega}{2T}}\right]
\label{4}
\end{eqnarray}
\end{widetext}
All integrals should be understood as principal parts.
 The terms $A(\omega)$ and $B(\omega)$ are non-singular at small frequencies and can be safely 
 approximated by $A(\omega) = A(0) = {\rm const}$, and $B(\omega) = B_1~ \omega$ ($B_1 = 1 + \lambda$ for vanishing $D(0)$).
 The non-trivial physics is due to the presence of the $C(\omega)$ term in
Eq.~(\ref{3}).  This term 
 originates from the imaginary part of the interaction 
$\chi ({\bf k} - {\bf k}^\prime, \Omega)$ along 
the real frequency axis and describes scattering by on-shell thermal bosons.
Neglecting $C(\omega)$  would lead to a conventional BCS-type result
$\Delta (\omega)  = A(\omega) \omega/B(\omega) =  A(0)/B_1 = 
{\rm const}$. 
However, 
the second term in the integral for 
$C(\omega)$ is  
proportional to $D(\omega) = \Delta (\omega)/\omega$, and
 diverges at zero frequency  if $\Delta (0)$ is nonzero. 
 Keeping only this term  
 and approximating the remainder of the integral by its $\omega\rightarrow 0$ limit,  we find from
 Eq.~(\ref{4}) $C (\omega)  = - i D(\omega) Q (T)$, where
\begin{equation}
Q(T)= \lambda \int_{-\infty}^\infty 
\frac{d \Omega}{\sinh{\frac{\Omega}{T}}} \frac{{\rm Im}\,\chi_s (\Omega)}{
\sqrt{1-D^2 (\Omega)}}
\label{6}
\end{equation} 
Upon substituting into Eq.~(\ref{3}) 
 we get 
 \begin{equation}
\Delta  (\omega)  =  \frac{A(0)}{B_1} ~
 \frac{\omega}{\omega + i \gamma},~~~ \gamma = \frac{Q(T)}{B_1}
\label{8}
\end{equation} 
This is a gapless superconductor -- the gap $\Delta$ 
is imaginary and linear in frequency at small frequencies, but recovers to a 
 real value, equal to the pairing amplitude $\Phi$, 
at  frequencies larger than $\gamma$. 
 We then obtain at small frequencies 
$Z (\omega) = Z_0 (1 + i \gamma/\omega)$,  
where $Z_0 = B_1\Phi/A_0$ is a mass-renormalization factor.
For the Green's function we then get
\begin{equation}
G (\omega, {\bf k}_F) = \frac{1}{Z_0}~
\frac{\omega + i \gamma}{(\omega + i\gamma)^2 - (\Phi/Z_0)^2 d^2_k}
\label{15}
\end{equation}  
This is precisely the $d$-wave single-lifetime model of Eq.~(\ref{1})
($\gamma_1 = \gamma_2$).  
Note that
 the gap amplitude $\Phi/Z_0$ and the
 scattering rate  
 $\gamma$ are  determined by different interactions.  
 In particular, $\gamma Z_0/\Phi=\gamma B_1/A_0 $ 
is strongly enhanced if the repulsive 
$s$-wave component of the interaction is much 
larger than the $d$-wave component.      

As stated earlier, the model of Eq.~(\ref{15}) 
gives rise to  arcs of the Fermi surface simply because the  condition 
for the arc, $ \gamma  > \sqrt{3} (\Phi/Z_0) |d_k|$,
 is always satisfied 
sufficiently close to the nodal direction, as long as $\gamma$ is non-zero, as it is for any non-zero $T$.

To understand the temperature dependence of the arc length, we note that 
 below the onset of pairing, $\Phi$ and $Z_0$ are weakly $T$-dependent, so that the 
 $T$ dependence of the arc length comes primarily from $\gamma (T)$.
  This $T$ dependence is given by Eq.~(\ref{6}), and
the functional form of $\gamma (T)$ depends on the form of $\chi_s (\Omega)$.  
As an example, consider the situation where well below $T^*$, 
${\rm Im} \chi_s (\Omega)$ can be approximated by an Einstein mode
 at some frequency $\Omega_{0}$, i.e.~${\rm Im}\, \chi_s (\Omega) = \pi
  \Omega^2_{0} ~\delta (\Omega^2 - \Omega^2_{0})$.  This can be a phonon or 
collective  spin excitations -- the latter form a gapless continuum in the normal state, but become a 
 mode (plus a gapped continuum) once $\Phi$ is nonzero. 

Substituting ${\rm Im}\, \chi_s (\Omega)$ into (\ref{4}) and (\ref{6}),
  we obtain that the boundary condition for the arc reduces to
\beq
\pi \eta~\frac{\lambda}{1+ \lambda} \frac{\Omega_{0}}{\sinh{\frac{\Omega_{0}}{T}}}
 = \sqrt{3} \frac{\Phi}{Z_0} |d_k|
\label{17}
\eeq
where $\eta$ is the normalized density of states at $\omega=\Omega_0$. For a circular Fermi surface, this is
 $\eta = (2/\pi) Re \int_0^{\pi/2}d \phi/\sqrt{1 - D^2 (\phi, \Omega_0)}$
 ($\eta >1$ if the largest contribution to $\eta$ comes from antinodal
 fermions).
The quantity $\Phi/Z_0$ is the measured gap at frequencies well above $\gamma$,
 and it is well approximated by the maximum value of the ARPES gap at the antinode.  We see from Eq.~(\ref{17}) 
that the arc length  is exponentially small at small $T \ll \Omega_{0}$, but becomes linear in $T$ at high temperatures.  This by itself is not a surprising result as at large $T$, the interaction is dominated by the thermal (zero-Matsubara frequency) term in which case the damping scales as
$T$ (the zero-Matsubara frequency contribution corresponds to approximating $\sinh(\Omega/T)$ in
Eq.~(\ref{6}) by $\Omega/T$).
A more interesting result, well known from the electron-phonon literature~\cite{allen}, is that, numerically, 
the r.h.s.~of Eq.~(\ref{17}) is essentially linear in $T$ already
from  $T \sim 0.3 \Omega_{0}$.

\begin{figure}
\centerline{\includegraphics[width=3.2in]{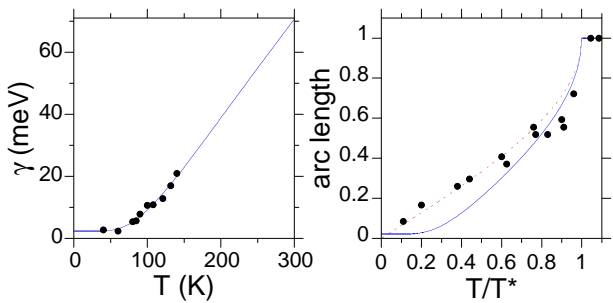}}
\caption{(a) Fit of $\gamma(T)$ as given by the left hand side of Eq.~(\ref{17}) to
values extracted from experimental ARPES data~\cite{prb01}.  A constant term of 2.3 meV has
been added to account for residual/instrumental
broadening at $T$=0.  The fitted value of $\Omega_0$ is 28 meV.
(b) Arc length versus $T$ from the fit in (a) assuming $(\Phi/Z_0)d_k =\Phi_0\cos(2\phi)$
with $\Phi_0=40$ meV and $T^*=300K$.
  Solid points are the data from Ref.~\onlinecite{arpes_camp}.
The dashed curve assumes instead that $\gamma \propto T$ for all $T$.}
\label{fig2}
\end{figure}

Experimentally, the arc length is linear in $T$ for $T_c < T < T^*$.
 This is roughly consistent with Eq.~(\ref{17}) if we associate 
$T_c$ with the mode energy (e.g.~for the neutron resonance mode~\cite{fong},
 $T_c \sim 0.2 \Omega_{0}$, in which case a linear in $T$ behavior sets in at  
 $T \geq 1.5 T_c$). 
   Moreover, this behavior will also
exhibit $T/T^*$ scaling as in Ref.~\onlinecite{arpes_camp} if $\Phi/Z_0$ scales with $T^*$,
as would be expected if $T^*$ were associated with pair formation.
To look into this further, we show
 in Fig.~2a  a plot of $\gamma(T)$ as given by the l.h.s.~of Eq.~(\ref{17})
versus the experimental $\gamma(T)$ values extracted from ARPES~\cite{prb01} 
on a slightly underdoped sample of Bi$_2$Sr$_2$CaCu$_2$O$_8$ ($T_c$=90K). 
  A good
fit is obtained with a reasonable value for $\Omega_0$ (28 meV) and a prefactor of one
(corresponding to 
 $\eta \lambda/(1+\lambda) =1$).  The resulting arc length is shown in Fig.~2b.
 The agreement is quite 
reasonable given the number of approximations made in deriving Eq.~(\ref{17}), and
the fact that we did not take into account
any temperature or doping dependence of $\Phi/Z_0$ and $\Omega_0$ when constructing these plots.
 We note that 
 a dependence of $\gamma$ that is strictly linear in $T$ is more consistent with the data, though
 a better fit can be obtained if $d_k$ is  flattened near the nodes as in Ref.~\onlinecite{varma}.
In addition, all the data shown in Fig.~2b were restricted to temperatures above $T_c$
(unlike in Fig.~2a).
  
In conclusion, we find that the temperature dependent
Fermi arcs observed by photoemission in the pseudogap phase can be explained
by lifetime broadening of a $d$-wave paired state.  We have investigated this in detail by examining
the effects of
inelastic scattering within the context of strong coupling Eliashberg theory, and find that the
results are in accord with the data. This suggests that $T^*$ can be thought of as an
energy scale associated with $d$-wave pair formation.

We acknowledge helpful discussions with J. C. Campuzano, I. Eremin, A. Kanigel, and 
M. Randeria, support
from NSF-DMR 0604406 (AVC) and DMR-0705847 (AJM), and
the U.S. DOE, Office of Science, under Contract 
No.~DE-AC02-06CH11357 (MRN).
This work was performed at the Aspen Center of Physics, and 
TU-Braunschweig (AVC).


\begin{thebibliography}{99}

\bibitem{arpes} D. S. Marshall \etal, Phys. Rev. Lett. {\bf 76}, 4841 (1996);
H. Ding \etal, Nature {\bf 382}, 51 (1996); A. G. Loeser \etal, Science {\bf 273}, 325 (1996).

\bibitem{norm_98} M. R. Norman \etal, Nature {\bf 392}, 157 (1998);
M. R. Norman \etal, Phys. Rev. B {\bf 57}, R11093 (1998).

\bibitem{arpes_camp} A. Kanigel \etal, Nature Physics {\bf 2}, 447 (2006).

\bibitem{latest} A. Kanigel \etal, arXiv:0708.4099.

\bibitem{varma} C. M. Varma and L. Zhu, Phys. Rev. Lett. {\bf 98}, 177004 (2007).

\bibitem{joerg} J. Schmalian, D. Pines and B. Stojkovic,  	
Phys. Rev. Lett. {\bf 80}, 3839 (1998).

\bibitem{tremblay} B. Kyung \etal, Phys. Rev. B {\bf 73}, 165114 (2006).

\bibitem{italy} A. Perali \etal,  Phys. Rev. B {\bf 62}, R9295 (2000). 

\bibitem{morr} S. Chakravarty \etal,  Phys. Rev. B {\bf 63}, 094503 (2001).

\bibitem{exception} An exception is the model of Ref.~\onlinecite{varma} that involves
an order parameter with $q=0$.  In this case, the arcs are due to lifetime broadening 
of a node  as in the present model 

\bibitem{gabi} K. Haule and G. Kotliar, arXiv:0709.0019.

\bibitem{phillips} T.D. Stanescu and P. Phillips, Phys. Rev. Lett. {\bf 91}, 017002 (2003).

\bibitem{Kotliar88} G. Kotliar and J. Liu, Phys. Rev. B {\bf 38}, 5142 (1988).

\bibitem{mohit} N. Trivedi and M. Randeria, Phys. Rev. Lett. {\bf 75}, 312 (1995).

\bibitem{Emery95} V. Emery and S. Kivelson, Nature {\bf 374}, 434 (1995).

\bibitem{millis_franz}  M. Franz and A. J. Millis, Phys. Rev. B {\bf 58}, 14572 (1998).

\bibitem{allen}  P. B. Allen, Comments Cond. Mat. Phys. {\bf 15}, 327 (1992).

\bibitem{mfl} C. M. Varma \etal, Phys. Rev. Lett. {\bf 63}, 1996 (1989).

\bibitem{timusk} see e.g. D. N. Basov and T. Timusk, Rev. Mod. Phys. {\bf 77}, 721 (2005).

\bibitem{elihu} E. Abrahams, M. Redi and J. W. F. Woo, Phys. Rev. B {\bf 1}, 208 (1970).

\bibitem{maki} C. Caroli and K. Maki, Phys. Rev. {\bf 159}, 316 (1967).

\bibitem{ag_gappl}  A. A. Abrikosov and L. P. Gor'kov, JETP {\bf 12}, 1243 (1961).

\bibitem{eliash} G.M. Eliashberg, Sov. Phys. JETP {\bf 11}, 696 (1960).

\bibitem{bergmann} G. Bergmann and D. Rainer, Z. Physik {\bf 263}, 59 (1973); J. P. Carbotte, Rev. Mod. Phys. {\bf 62}, 1027 (1990).

\bibitem{msv} A. J. Millis, S. Sachdev and C. M.  Varma, Phys. Rev. B {\bf 37}, 4975 (1988).

\bibitem{ssw} D. J. Scalapino, Y. Wada and J. C. Swihart, Phys. Rev. Lett. {\bf 14}, 102 (1965). 

\bibitem{comb}  R. Combescot, Phys. Rev. B {\bf 51}, 11625 (1995);
A. E. Karakozov, E. G. Maksimov and A.A. Mikhailovsky, Solid State  Comm. {\bf 79}, 329 (1991);
F. Marsiglio, M. Schossmann and J. P. Carbotte, Phys. Rev. B {\bf 37}, 4965 (1988).

\bibitem{acf} Ar. Abanov, A. Chubukov and A. Finkelstein, Europhys. Lett.
 {\bf 54}, 488 (1999);  
A. Abanov \etal, (unpublished).

\bibitem{prb01}
M. R. Norman \etal, Phys. Rev. B {\bf 63}, 140508 (2001).

\bibitem{fong} H.F. Fong \etal, Nature {\bf 398}, 588 (1999).

\end{thebibliography}
 \end{document}